\documentclass[acmtog, nonacm]{acmart}
\AtBeginDocument{%
  \providecommand\BibTeX{{%
    \normalfont B\kern-0.5em{\scshape i\kern-0.25em b}\kern-0.8em\TeX}}}

\usepackage{soul}
\usepackage{multirow}
\newcommand{\revision}[1]{#1}

\newcommand{\ie}{\emph{i.e.}}
\newcommand{\eg}{\emph{e.g.}}

\usepackage{enumitem}

\copyrightyear{2024} 
\acmYear{2024} 
\setcopyright{rightsretained} 
\acmConference[SA Conference Papers '24]{SIGGRAPH Asia 2024 Conference Papers}{December 3--6, 2024}{Tokyo, Japan}
\acmBooktitle{SIGGRAPH Asia 2024 Conference Papers (SA Conference Papers '24), December 3--6, 2024, Tokyo, Japan}\acmDOI{10.1145/3680528.3687562}
\acmISBN{979-8-4007-1131-2/24/12}

\begin{document}

\title{Dance-to-Music Generation with Encoder-based Textual Inversion}


\author{Sifei Li}
\orcid{0009-0003-9331-4593}
\affiliation{
\institution{MAIS, Institute of Automation, CAS}
\country{China}
}
\affiliation{
\institution{School of Artificial Intelligence, UCAS}
\country{China}
}
\email{lisifei2022@ia.ac.cn}

\author{Weiming Dong}
\authornote{Corresponding author: Weiming Dong (weiming.dong@ia.ac.cn).}
\orcid{0000-0001-6502-145X}
\affiliation{
\institution{MAIS, Institute of Automation, CAS}
\country{China}
}
\affiliation{
\institution{School of Artificial Intelligence, UCAS}
\country{China}
}
\email{weiming.dong@ia.ac.cn}

\author{Yuxin Zhang}
\orcid{0000-0001-6433-2678}
\affiliation{
\institution{MAIS, Institute of Automation, CAS}
\country{China}
}
\affiliation{
\institution{School of Artificial Intelligence, UCAS}
\country{China}
}
\email{zhangyuxin2020@ia.ac.cn}

\author{Fan Tang}
\orcid{0000-0002-3975-2483}
\affiliation{
\institution{University of Chinese Academy of Sciences}
\country{China}
}
\email{tfan.108@gmail.com}
 
\author{Chongyang Ma}
\orcid{0000-0002-8243-9513}
\affiliation{
\institution{Kuaishou Technology}
\country{China}
}
\email{chongyangm@gmail.com}

\author{Oliver Deussen}
\orcid{0000-0001-5803-2185}
\affiliation{
\institution{University of Konstanz}
\country{Germany}
}
\email{oliver.deussen@uni-konstanz.de}

\author{Tong-Yee Lee}
\orcid{0000-0001-6699-2944}
\affiliation{
\institution{National Cheng-Kung University}
\country{Taiwan}
}
\email{tonylee@ncku.edu.tw}

\author{Changsheng Xu}
\orcid{0000-0001-8343-9665}
\affiliation{
\institution{MAIS, Institute of Automation, CAS}
\country{China}
}
\affiliation{
\institution{School of Artificial Intelligence, UCAS}
\country{China}
}
\email{csxu@nlpr.ia.ac.cn}

\renewcommand{\shortauthors}{Li, et al.}

\begin{teaserfigure}
\centering
   \includegraphics[width=\linewidth]{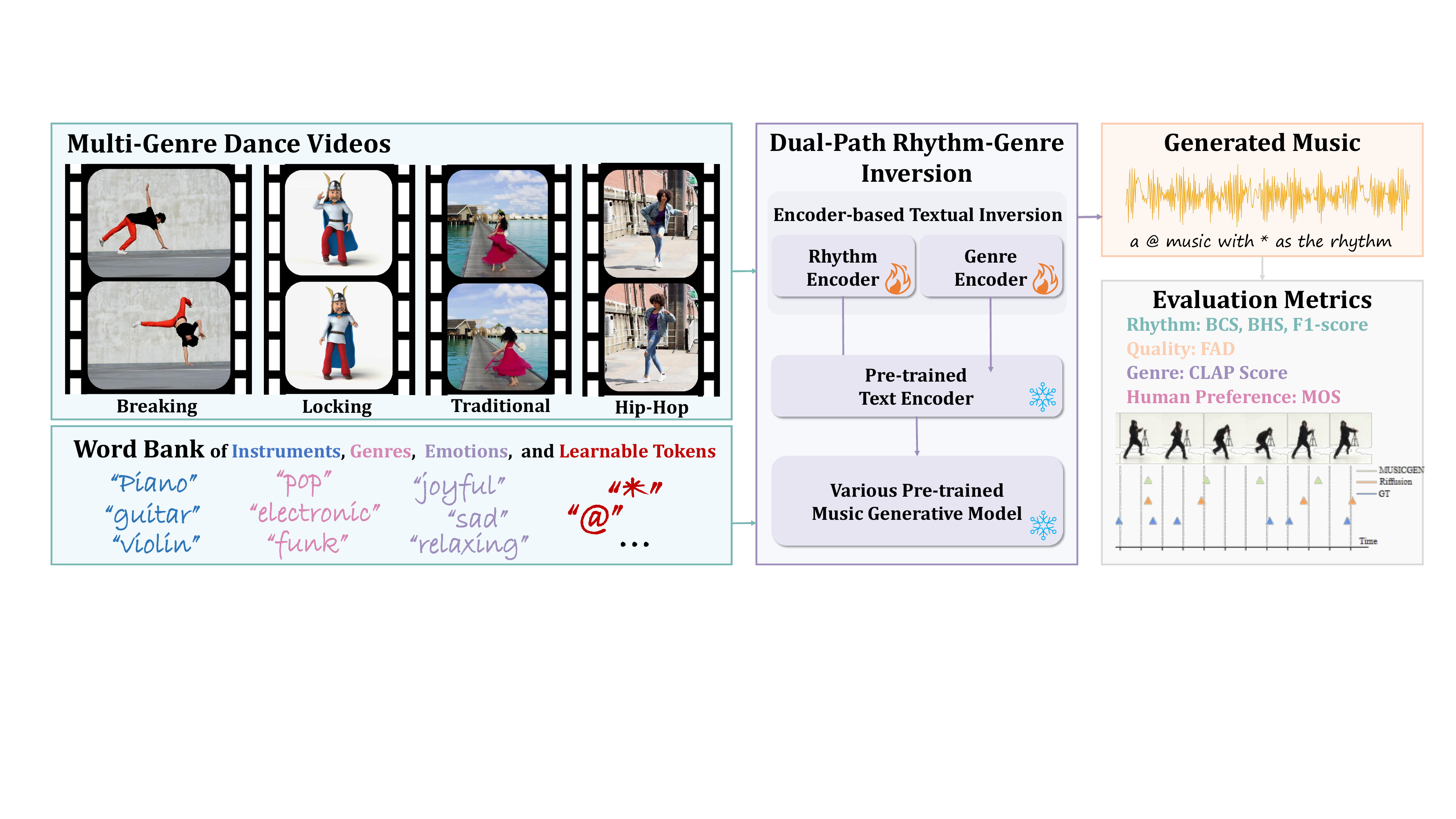}
   \caption{
   We propose dual-path rhythm-genre inversion to incorporate rhythm and genre information from a dance motion sequence into two learnable tokens, which are then used to enhance the pre-trained text-to-music models with visual control. 
   Through encoder-based textual inversion, our method offers a plug-and-play solution for text-to-music generation models, enabling seamless integration of visual cues. The word bank comprises learnable pseudo-words (represented as ``@'' for genre and ``*'' for rhythm) and descriptive music terms such as instruments, genres, and emotions.
   By combining these pseudo-words and descriptive music terms, our method allows for the generation of a diverse range of music that is synchronized with the dance. 
   }
\label{fig:teaser}
\end{teaserfigure}



\begin{abstract}
The seamless integration of music with dance movements is essential for communicating the artistic intent of a dance piece.
This alignment also significantly improves the immersive quality of gaming experiences and animation productions. Although there has been remarkable advancement in creating high-fidelity music from textual descriptions, current methodologies mainly focus on modulating overall characteristics such as genre and emotional tone. They often overlook the nuanced management of temporal rhythm, which is indispensable in crafting music for dance, since it intricately aligns the musical beats with the dancers' movements.
Recognizing this gap, we propose an encoder-based textual inversion technique to augment text-to-music models with visual control, facilitating personalized music generation.
Specifically, we develop dual-path rhythm-genre inversion to effectively integrate the rhythm and genre of a dance motion sequence into the textual space of a text-to-music model.
Contrary to traditional textual inversion methods, which directly update text embeddings to reconstruct a single target object, our approach utilizes separate rhythm and genre encoders to obtain text embeddings for two pseudo-words, adapting to the varying rhythms and genres. 
We collect a new dataset called In-the-wild Dance Videos (InDV) and demonstrate that our approach outperforms state-of-the-art methods across multiple evaluation metrics. Furthermore, our method is able to adapt to changes in tempo and effectively integrates with the inherent text-guided generation capability of the pre-trained model.
Our source code and demo videos are available at \url{https://github.com/lsfhuihuiff/Dance-to-music_Siggraph_Asia_2024}.
\end{abstract}
\maketitle

\section{Introducion}

If life is a sort of dance, then music is the enchanting melody that accompanies our every graceful movement. Music's power lies in its ability to synchronize and enhance visual narratives, a trait extensively utilized in the gaming and film industries. The advent of short video content has seen a surge in amateur bloggers sharing dance videos on social media, showcasing music's integral role in this domain.
Classical approaches~\cite{lee2005automatic,davis2018visual} align audio and video by time-warping the video. However, this process often necessitates the manual selection of suitable music and introduces variations to the original video. Additionally, copyright issues related to music can impede the widespread distribution of videos. Consequently, creating an original soundtrack for a video remains a challenging task.

As AI-generated content continues to advance, there has been a surge in research focusing on dance-to-music generation, aligning with contemporary trends and reflecting the broader evolution of AI applications in creative domains.
Some approaches~\cite{gan2020foley, su2021does, di2021video, aggarwal2021dance2music, han2023dance2midi} employ symbolic music representation, such as MIDI and REMI, to generate music for dance videos. However, the generated music is often limited to a few classical instruments (\eg, piano, guitar). Furthermore, these methods often result in incoherent note sequences and simplistic melodies, deviating significantly from the characteristics of real dance music.
Other approaches~\cite{zhu2022quantized, zhu2022discrete,yu2023long, tan2023motion} directly use waveforms or mel-spectrograms to generate music for dance videos, resulting in more intricate and nuanced musical compositions. However, the generated audio quality is often suboptimal, and some tend to overemphasize percussive elements such as drum beats, failing to capture melodic nuances adequately.
Moreover, these approaches are inherently limited by the musical genres prevalent in the training dataset, making it challenging to generate diverse and suitable music for dance videos across various styles.
Consequently, their real-world applicability remains constrained.
Pre-trained text-to-music models~\cite{Forsgren_Martiros_2022,copet2023simple}  demonstrate their ability to generate high-quality and diverse music. However, most current models only provide control over global attributes of the music and cannot manipulate local properties such as rhythm.

To address the above-mentioned problems, we propose a novel approach for dance-to-music generation that leverages arbitrary pre-trained text-to-music models.
Our method seamlessly integrates rhythm information extracted from dance videos into these models, while preserving their original text-based generation capabilities.
Given an input dance video, the model generates synchronized music that harmoniously complements the dance movements.
As illustrated in Fig.~\ref{fig:teaser}, our method facilitates dance-to-music generation across various genres (\eg, ``pop'', ``lock'', ``house''). Additionally, by leveraging textual descriptions, users can exert control over high-level musical attributes like genre and emotion.
Notably, our approach's reliance on motion sequences for rhythm extraction allows for its extension to diverse physical activities (\eg, ``jump rope'', ``artistic gymnastics'').

To achieve this goal, we develop a method to incorporate rhythm information into a pre-trained text-to-music model.
Inspired by Textual Inversion~\cite{gal2023TI} which employs a pseudo-word to represent a specific concept through image reconstruction, we aim to learn a pseudo-word that represents rhythm information inherent in arbitrary dance videos and integrate it into the vocabulary of the text encoder during training.
However, in the text encoder, the text embeddings corresponding to an individual token are fixed, while the rhythms in different videos are variable, making it challenging to create a pseudo-word for each rhythm pattern.
Therefore, in contrast to prior methods~\cite{gal2023TI, zhang2023inversion, huang2023reversion, mokady2023null} using a fixed text embedding for each pseudo-word, we propose an encoder-based textual inversion method that incorporates an encoder branch that enables the text embedding associated with the pseudo-word to dynamically adapt to varying input conditions.
As a result, our method enables the generation of music with different rhythms based on different motion sequences in the video.

To narrow the gap between motion sequences and text embeddings, we employ a hybrid approach that combines traditional feature extraction techniques, extracting rhythm sequences from the motion sequences, and projector networks to construct the rhythm encoder. Furthermore, we introduce a genre encoder to provide additional control over music generation.

Our contributions can be summarized as follows:
\begin{itemize}[leftmargin=*]
    \item We propose a novel encoder-based textual inversion method that enables a single pseudo-word to represent a class of variable attributes rather than a fixed object.
    \item We develop rhythm and genre encoders to achieve dual-path rhythm-genre inversion, converting rhythm and genre information into text embeddings. This enhances text-to-music models by integrating visual control. In dance-to-music generation, it provides flexibility in high-level feature control of music with text.
    \item  We collect a challenging dance-music dataset entitled ``In-the-wild Dance Videos'' (InDV).
    Unlike AIST++, this dataset encompasses a wide array of dance movements within individual videos and incorporates dance genres characterized by intricate rhythmic structures, such as Chinese traditional dance.
    \item 
    Experimental results indicate that our method attains superior performance across various evaluation metrics. Additionally, our approach demonstrates robust adaptability to tempo variations and in-the-wild data.
   
\end{itemize}

\section{Related Work}
\paragraph{Audio-video synchronization}

Some works~\cite{lee2005automatic, davis2018visual, sun2023eventfulness} achieve audio-video synchronization through editing techniques.
Among them, \cite{davis2018visual} and \cite{sun2023eventfulness} enable alignment between audio and video and propose methods for quantifying video rhythm.
Specifically, \citet{davis2018visual} calculate a directogram using optical flow to detect visible impacts, while \citet{sun2023eventfulness} learn visual eventfulness to capture video rhythm.
While these works focus on editing methods, our approach extracts video rhythm from 2D keypoints and embeds it into a pre-trained model to generate new music that synchronizes with the videos.

Generating audios synchronized with input videos~\cite{jin2022neuralsound, qi2023rd, du2023conditional, su2023physics} has become increasingly popular in recent years.
Some works~\cite{su2020audeo, gan2020foley, su2021does} use symbolic representation for video music generation.
Controllable Music Transformer (CMT)~\cite{di2021video} designs rule-based rhythmic relationships, allowing control over timing, motion, beat, and music genre for background music generation.
DANCE2MIDI~\cite{han2023dance2midi} constructs the D2MIDI dataset for multi-instrument MIDI and dance pairing, enabling the generation of coherent music sequences from dance videos.  
Video2Music~\cite{kang2023video2music} and V-MusProd~\cite{zhuo2023video} leverage multiple video features and transformer models to generate music that is synchronized with videos.
However, symbolic representation-based methods overlook timbre and dynamics, limiting musical expressiveness and variation in the generated music.

Recent studies have investigated the use of spectrograms or audio waveforms for video-to-music generation.
Dance2Music-GAN (D2M-GAN)~\cite{zhu2022quantized} is an adversarial multi-modal framework that generates complex music samples conditioned on dance videos. 
\citet{zhu2022discrete} introduce a novel approach that maximizes mutual information using a conditional discrete contrastive diffusion (CDCD) loss to generate dance music. 
LORIS~\cite{yu2023long} employs a latent conditional diffusion probabilistic model and context-aware conditioning encoders for synthesizing long-term conditional waveforms.
\citet{tan2023motion} combines a UNET-based latent diffusion model and a pre-trained VAE model to generate plausible dance music from 3D motion data and genre labels. 
However, most of them generate low-quality audio with noticeable noise, and the generated music tends to be similar, making it challenging to adapt to diverse video scenarios in the wild. In contrast, our approach, based on a pre-trained text-to-music model, is capable of generating diverse and high-quality beat-aligned dance music.
V2Meow~\cite{su2023v2meow} utilizes a multi-stage autoregressive model to generate visually-aligned music, allowing for high-level feature control with text. However, it requires training from scratch on a large-scale O(100k) dataset, in contrast to our approach, which employs a small-scale O(1k) dataset. By leveraging encoder-based textual inversion, we drive a pre-trained text-to-music model to achieve dance-to-music generation.

\paragraph{Text-to-music generation}
Text-to-music generation has made remarkable advances in recent years. Any-to-Any generation such as CoDi~\cite{tang2023any} and NExT-GPT~\cite{wu2023next} have the capability to generate across different modalities, including text-to-music generation. 
LLARK~\cite{gardner2023llark} combines a pre-trained music generative model with a pre-trained language model for music understanding.
Several approaches like Make-an-Audio~\cite{huang2023make}, AudioLDM~\cite{liu2023audioldm}, AudioLDM2~\cite{liu2023audioldm2}, and Tango~\cite{ghosal2023text} utilize diffusion models to generate text-guided audio encompassing speech, sounds, and music. However, these methods face limitations in terms of data quality and model scalability, resulting in lower-quality music generation. Some works such as Archisound~\cite{schneider2023archisound}, Riffusion~\cite{Forsgren_Martiros_2022}, Mo\^{u}sai~\cite{schneider2023mo}, and Noise2Music~\cite{huang2023noise2music} leverage diffusion models to generate high quality music from textual input.
Furthermore, transformer-based frameworks like MusicLM~\cite{agostinelli2023musiclm}, MUSICGEN~\cite{copet2023simple}, and MuseCoco~\cite{lu2023musecoco} encode music as discrete tokens, enabling the generation of high-quality music with rich textual description.
Mustango~\cite{melechovsky2023mustango} enhances text-to-music models with control over harmony, rhythm, and dynamics through a text format. However, these methods are limited to the single modality of music and primarily manipulate high-level features, failing to establish a connection between visual features and music.
In contrast, our method employs learnable pseudo-words to achieve low-level control, enabling the generation of music that aligns with video content.

\paragraph{Personalization of generative models}
While text-guided content generation has achieved impressive results, relying solely on text is insufficient for precise control over the generated content, especially when targeting specific objectives. Therefore, the personalization of generative models has become a recent research focus. Personalization of generative models refers to the generation of content personalized to specific objects based on text-to-content models, leveraging prompt learning or fine-tuning techniques.
DreamBooth~\cite{ruiz2023dreambooth} introduces a class-specific prior preservation loss for personalized image generation.
LORA~\cite{hu2021lora} proposes an efficient fine-tuning method known for achieving remarkable performance in personalized generation.
EDICT~\cite{wallace2023edict} proposes an exact diffusion inversion method for image editing, which requires no model training/finetuning, prompt tuning, or extra data.
\citet{gal2023TI} introduce a textual inversion (TI) method that progressively updates the embeddings of text placeholders corresponding to specific object's visual features within a pre-trained text encoder.
Inspried by TI, many variants~\cite{gal2023encoder, huang2023reversion, li2023stylediffusion, voynov2023p+, mokady2023null, zhang2023inversion, zhang2023prospect, Alaluf:2023:NST} achieve high-quality and more controllable personalized image generation.
\citet{li2024music}, \citet{plitsis2023investigating}, \citet{novack2024ditto}, \citet{manor2024zero} 
as well as \citet{wu2023music} (MusicControlNet) 
explore personalized music generation using pre-trained models. However, none of these methods successfully established a connection between visual features and music.
In dance-to-music generation, the incorporation of rhythm information from dance videos is necessary.
Thus, as a new task of TI, we propose an encoder-based textual inversion approach that facilitates personalized music generation, allowing for the generation of music that is aligned with specific rhythmic video.

\section{Method}
\begin{figure*}[ht]
\centering
   \includegraphics[width=1\linewidth]{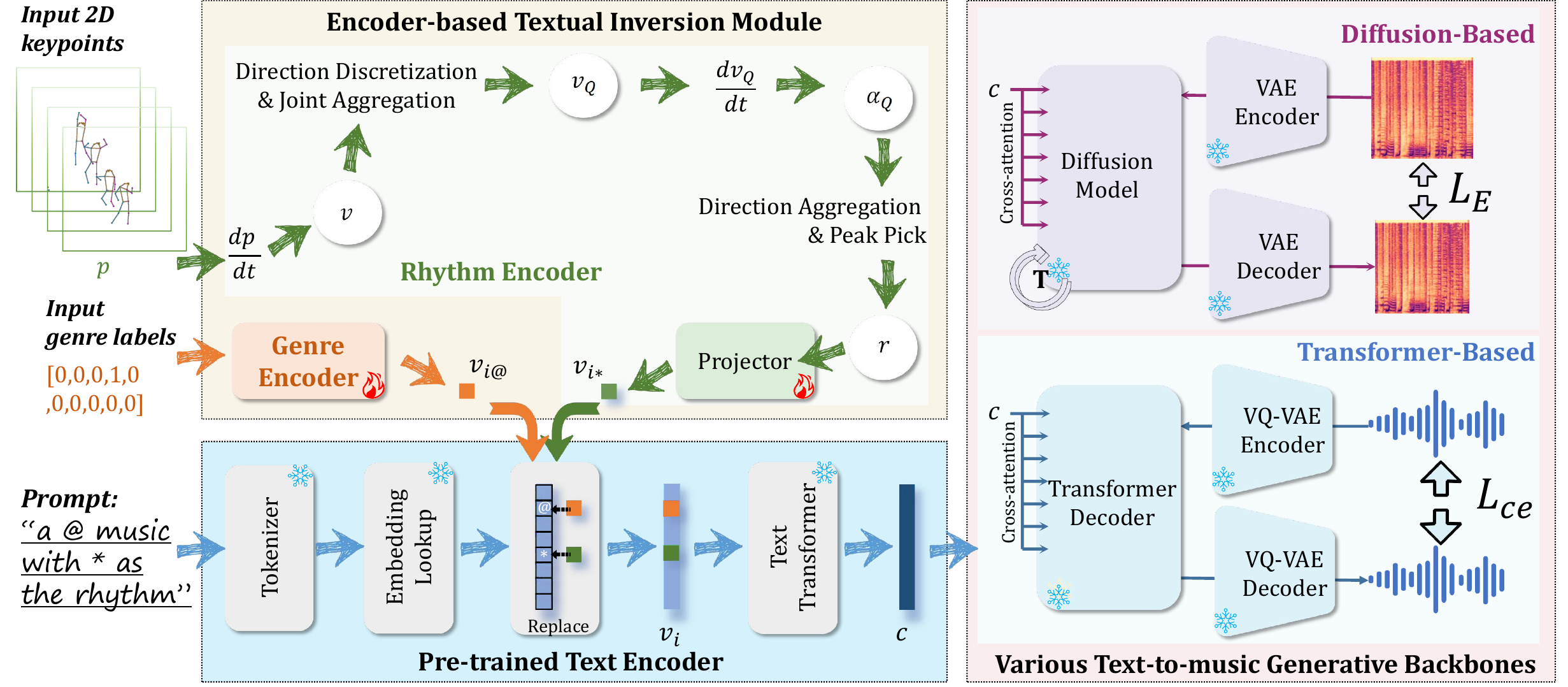}
   \caption{
   We employ various pre-trained music generative models as the generative backbone and propose an encoder-based textual inversion method. During training, we fix the prompt as ``a @ music with * as the rhythm'', where ``@'' and ``*'' respectively represent the placeholders for the genre and rhythm of the input dance. Our dual-path rhythm-genre inversion optimizes the rhythm encoder and genre encoder together during training.
   Parameters $v_i$, $v_{i@}$, and $v_{i*}$ correspond to the text embeddings of the prompt, ``@'', and ``*'', respectively. }
\label{fig:method}
\end{figure*}
In this paper, we propose an encoder-based textual inversion technique, implemented as a separate network structure, that seamlessly integrates with text-to-music generation models.
They are augmented with our dual-path rhythm-genre inversion, utilizing learnable pseudo-words that integrate rhythm and genre information to guide the dance-to-music generation.
During training, the prompt remains fixed as ``a @ music with * as the rhythm'', where ``@'' represents the genre description and ``*'' indicates the rhythm description. We obtain text embeddings for two placeholder words using separate rhythm and genre encoders, resulting in two controllable tokens. By reconstructing the target audio, we simultaneously optimize the parameters of both encoders, facilitating their mutual enhancement.
In the inference phase, flexible text descriptions can be used in addition to the fixed prompt to exert control over the high-level feature of music. The inclusion of the pseudo-word ``*'' representing rhythm is vital to ensure the generation of music that aligns with the dance.

\subsection{Encoder-based Textual Inversion}
Our objective is to leverage the generative capabilities of a pre-trained text-to-music model to generate music for dance videos. Typically, a text encoder firstly tokenizes a prompt into multiple indices, each corresponding to an embedding in the corresponding embedding lookup. 
A text transformer then encodes these text embeddings to serve as conditional guidance for the generative model. 
We employ three prominent music generation frameworks as backbones: Riffusion~\cite{Forsgren_Martiros_2022}, AudioLDM~\cite{liu2023audioldm}, and MUSICGEN~\cite{copet2023simple}.
For Riffusion and AudioLDM, the generative models employed are the Latent Diffusion Models (LDMs)~\cite{latentdiffusion}, with the text encoders being the text encoder of CLIP \cite{radford2021learning} and CLAP \cite{wu2023large}, respectively.
For MUSICGEN, the text encoder utilized is the T5~\cite{raffel2020exploring}, while the generative model is the Transformer decoder. 

A prevalent issue in most text-to-music models is their limited ability to control local attributes of music (\eg, rhythm), due to the difficulty of accurately describing these attributes in natural language.
This limitation poses a significant challenge when generating music that aligns with dance videos, where precise control over rhythm is essential.
Textual inversion~\cite{gal2023TI}, a personalization method for image generation, employs a pseudo-word (\eg, ``*'') as a placeholder for a specific object and iteratively optimizes the text embedding associated with the pseudo-word. This process aims to integrate the textual description of the specific object into the vocabulary of the pre-trained text encoder, thereby enabling targeted editing of the specific object. The optimization objective can be defined as follows:
\begin{equation}
    v_{i*}=\underset{v}{\arg\min}\,\mathbb{E}_{z,y,\epsilon,t}[\|\epsilon-\epsilon_{\theta}(z_{t}, t, c_{\theta}(y))\|_{2}^{2}],
\end{equation}
where $z \sim E(x), \epsilon \sim \mathcal{N}(0,1)$, $\epsilon_{\theta}$ denotes LDMs, $c_{\theta}$ denotes the text encoder of CLIP, and $y$ denotes the prompt.

It is natural to consider using a pseudo-word to represent the rhythm of a video and guide the generation of music, \eg, ``lively melody with * as the rhythm and easy instrumental arrangement''. However, textual inversion requires training a separate model for each object, with fixed embeddings for all pseudo-words.
Yet, the rhythm of dance is highly variable, and training a model for the rhythm of each video would be costly. Therefore, we propose an encoder-based textual inversion method, which allows us to define the rhythm as a variable attribute in inversion.

To enhance the expressiveness of individual pseudo-words, we introduce an encoder branch into the text encoder. This encoder extracts features from a specific input type (\eg, human poses), and maps them to the textual space. This enables a single pseudo-word with the capacity to exhibit distinct text embeddings contingent upon the input, allowing for control over the generation of diverse content. The encoder is iteratively optimized by reconstructing the target objects. For Riffusion and AudioLDM, the loss function of the encoder is defined as follows:
\begin{equation}
    L_E=\mathbb{E}_{z,x,y,\epsilon,t}[\|\epsilon-\epsilon_{\theta}(z_{t}, t, t_{\theta}(x,y))\|_{2}^{2}],
\end{equation}
where $x$ denotes a specific type of input, $t_{\theta}$ denotes the expanded text encoder.
For MUSICGEN, the cross-entropy loss function of the encoder is defined as follows:
\begin{equation}
    L_{\text{ce}} = \frac{1}{K} \sum_{k=1}^{K} \text{CE}(G_{\theta}(t_{\theta}(x,y))_k, {t}_k),
\end{equation}
where $t$ denotes target tokens, $x$ denotes a specific type of input,  $t_{\theta}$ denotes the expanded text encoder, $G_{\theta}$ denotes the generative model and $K$ denotes the index of codebook.

Utilizing a single pseudo-word enables precise control over specific attributes of the generated content (\eg, rhythm). This control can seamlessly integrate with human text editing, resulting in enhanced flexibility. 
The encoder-based textual inversion framework is not confined to a specific task but can be applied to different areas, where the attributes are variable and challenging to describe using natural language. In this study, we demonstrate our approach through dual-path rhythm-genre inversion, employing rhythm and genre encoders to facilitate dance-to-music generation.

\subsection{Rhythm Encoder}

We combine traditional feature extraction with a projector. ``Dance to the rhythm'' serves as an important connection point for synchronizing music and video. Conversely, ``drop the beat to the motion'' is crucial for the task of creating music for dance videos. Dancers typically perform actions or transitions in sync with specific musical beats during dancing. The initiation and transitions of movements align with points of local maximum acceleration in kinematics. Based on LORIS~\cite{yu2023long}, we calculate the first-order difference of the 2D keypoints $p(t,j,c)$ of the dancers over time to obtain the motion velocities, where $j$ represents joints and $c$ denotes coordinates. We use $v_x$ and $v_y$ to represent the velocities in the x and y directions, respectively.
Subsequently, we discretize them
into $K$ intervals, a process which we refer to as direction discretization:
\begin{equation}
    v_Q(t,j,k)=Vl_{\theta}(t,j,k),
\end{equation}
\begin{equation}
    l_{\theta}(t,j,k) = \begin{cases} 
      1, & \text{if } \text{$k=[\frac{\theta}{{2\pi}/{K}}]$} \\
      0, &  \text{otherwise} \\
   \end{cases}
\end{equation}
\begin{equation}
    V = \sqrt{{v_x}^2 + {v_y}^2},\\
    \theta = \arctan\left(\frac{v_y}{v_x}\right), 
\end{equation}
where $\theta$ represents the direction angle, and $V$ represents the magnitude of the velocity.
By calculating the first-order difference in the temporal dimension of $v_Q$ and aggregating the acceleration values across bins, we obtain the discrete acceleration $a_Q$.

We retain the positive acceleration values and compute the sum of these accelerations across joint and directional dimensions to obtain the total acceleration, denoted as $a$:
\begin{equation}
    a = \sum_{j,k} a_Q(t,j,k).
\end{equation}
Next, we determine the local maxima of $a$ within a given time window. Subsequently, we construct a rhythm sequence $r$, where positions corresponding to the local maxima are set to 1, while all other positions are set to 0.
We then design a projector to map $r$ to the textual space, resulting in an embedding denoted as $v_{i*}$, which replaces the text embedding for the pseudo-word ``*''.

\subsection{Genre Encoder}

Reconstructing the audio mel-spectrogram solely from rhythmic information may introduce music genre information into the rhythmic pseudo-word, which hinders attribute disentanglement. Furthermore, it also results in a lack of control in music generation based on the dance genre. To overcome this, we introduce a genre encoder to enhance the generative model's reconstruction capability by incorporating genre information. We represent dance genres using one-hot encoding. By employing a combination of linear and activation layers, we map the one-hot encoding to the textual space, effectively replacing the text embedding associated with the pseudo-word ``@'' with the corresponding genre embedding $v_{i@}$.

\section{Experiments}

\begin{table*}
   \caption{Quantitative evaluation and user study results in comparison with state-of-the-arts methods on AIST++ dataset.
   The best results are in highlighted \textbf{bold} and the second best ones are \underline{underlined} (same in the following tables).}
   \centering
   \begin{tabular}{c||cccc|c|c|cc|c}
   \hline
  {}&\multicolumn{4}{|c}{Rhythm} & \multicolumn{1}{|c}{Quality} & \multicolumn{1}{|c}{Genre } & \multicolumn{2}{|c}{MOS} &\multicolumn{1}{|c}{Inference Time}\\
  \cline{2-10}
     & BCS $\uparrow$  & BHS $\uparrow$ & F1-score  $\uparrow$ & TD $\downarrow$  & FAD $\downarrow$ & CLAP $\uparrow$ & Coherence $\uparrow$ & Quality $\uparrow$ & s/clip $\downarrow$\\ 
   \hline
   Ground Truth & -&-&-&-&- &- &4.26&4.25 &-\\
   CMT~\cite{di2021video} & 0.3368  & 0.1515  & 0.2090 &21.74  & 16.54 & 0.4454 & 1.79 & 3.08 & 6.44\\
   CDCD~\cite{zhu2022discrete} & 0.4233 & 0.2151  & 0.2852 &19.25 & 16.47 & 0.3032 & 2.02 & 1.35 & \underline{5.72}\\
   LORIS~\cite{yu2023long} & 0.3721 & 0.3371  & 0.3537 &\underline{17.80}  & 13.15 & 0.6180 & 2.56 & 2.42 &21.2\\
   MDM~\cite{tan2023motion} & 0.3798 &  \underline{0.4185}  & 0.3982 &22.96  & \underline{4.812} & 0.5793 & 2.97 & 2.58 & \textbf{3.71}\\
  \textbf{Ours (AudioLDM)} & \underline{0.4419} & 0.3605 & 0.3971 &22.73  & 8.522 & \underline{0.7030} &2.95 & \textbf3.02 & 14.82\\
   \textbf{Ours (MUSICGEN)} & 0.4118 & 0.3874 &  \underline{0.3992} & \textbf{16.06}  & 6.014 & 0.4685 &\textbf{3.56} & \textbf{3.54} & 11.48\\
   \textbf{Ours (Riffusion)}& \textbf{0.4761}  & \textbf{0.4398} & \textbf{0.4572} & 20.34  & \textbf{3.416} & \textbf{0.7680} & \underline{3.24} & \underline{3.15} & 8.49 \\
   \hline
   \end{tabular}
   \label{tab:contrast}
\end{table*}

\begin{table}
   \caption{Comparision results of the phase-aligned version of each output. OA, OM, and OR represent Ours (AudioLDM), Ours (MUSICGEN), and Ours (Riffusion), respectively.}
   \centering
   \resizebox{0.96\linewidth}{!}{
   \begin{tabular}{c||cccc|c|c}
   \hline
  {}&\multicolumn{4}{|c}{Rhythm} & \multicolumn{1}{|c}{Quality} & \multicolumn{1}{|c}{Genre} \\
  \cline{2-7}
     & BCS $\uparrow$  & BHS $\uparrow$ & F1-score  $\uparrow$ & TD $\downarrow$  & FAD $\downarrow$ & CLAP $\uparrow$ \\ 
   \hline
   CMT & 0.4038  & 0.1850  & 0.2538 & 21.74 & 16.94 & 0.5135 \\
   CDCD & \textbf{0.4619} & 0.2318  & 0.3087& 21.66 &16.71 & 0.2845\\
   LORIS & \underline{0.4476} & 0.3438  & 0.3889 & 21.39 &13.42 & 0.6216 \\
   MDM & 0.3460 &  \underline{0.4270}  & 0.3823 & 21.44 &\underline{4.258} & 0.5889  \\
   \textbf{OA} & 0.4073 & 0.3351 &  0.3677& 26.40 &8.739 & \underline{0.6853} \\
   \textbf{OM} & 0.4368 & \textbf{0.4341} &  \textbf{0.4354} & \textbf{15.19} &5.762 & 0.6384\\
   \textbf{OR}& 0.4452 & 0.4251 & \underline{0.4349} & \underline{20.34}& \textbf{3.201} & \textbf{0.7453} \\
   \hline
   \end{tabular}
   }
   \label{tab:contrast_align_phase}
\end{table}

\subsection{Experimental Setup}

\paragraph{Dataset}
We evaluate our method on two dance-music datasets, \ie, AIST++~\cite{Li:2021:AIST} and our newly collected InDV dataset, both consisting of ten genres.
The AIST++ dataset comprises 70 songs paired with 460 Choreographies, evenly distributed across 10 different dance genres, along with corresponding 2D keypoints. We employ the official training, validation, and testing sets
and segment the data into 5.12-second clips. As a result, our final training, validation and testing sets consist of 2744, 36, and 36 samples, respectively. It is free for research purpose and we use this dataset for the main experiments and evaluations.
We also collect and annotate a new dance-music dataset called InDV (In-the-wild Dance Videos), which contains 595 5.12-second clips with 216 songs. The dataset is categorized into 10 genres:  Ballet, Breaking, Chinese Traditional Dance, Hip-hop, Jazz, Latin, Locking, Popping, and Waacking. The training, validation, and testing sets consist of 544, 25, and 26 samples, respectively.
We employ HRNet~\cite{sun2019deep} in mmPose~\cite{mmpose2020} to obtain 2D skeletons (dance motion sequences) from in-the-wild data.
The video fps for the keypoints is 60.

\paragraph{Implementation details}
We conduct experiments on our encoder-based textual inversion method using three prominent music generation frameworks: diffusion-based models Riffusion~\cite{Forsgren_Martiros_2022} and AudioLDM~\cite{liu2023audioldm}, and autoregressive model MUSICGEN~\cite{copet2023simple}.
For Riffusion, 
we use the default hyperparameters of LDMs and employ a base learning rate of 0.0005. The model training is executed on an NVIDIA GeForce RTX 3090 GPU, taking approximately 12 hours to complete over 50 epochs.
As for MUSICGEN,
we apply the AdamW optimizer with $\beta_1 = 0.9, \beta_2 = 0.95$, and a weight decay of $0.01$. A warm-up learning rate of 1e-4 is used for all layers during the initial $6000$ training iterations.
We train the network for 20 epochs on a single NVIDIA A40 GPU, taking approximately 10 hours.
For AudioLDM, we keep the default hyperparameters and train it for 50 epoch on a single NVIDIA A40 GPU, taking approximately 11 hours.

\subsection{Evaluation Metrics}
Inspired by previous work and further analysis of the task, we develop a comprehensive evaluation protocol that incorporates multiple metrics to evaluate from the following perspectives.

\paragraph{Rhythm}
Regarding the musical beats, a tolerance offset of one second is easily discernible to the human ear. To enhance the accuracy of evaluation, we adjust the tolerance offset from 1 second to 0.2 second. 
Specifically, we follow the setup of LORIS~\cite{yu2023long}, where $B_g$ denotes the number of musical beats in the generated music, $B_t$ denotes the number of musical beats in the ground-truth music, and $B_a$ denotes the number of aligned musical beats. We also conduct evaluation on global rhythm metrics. The evaluation metrics related to rhythm alignment are as follows:
\begin{itemize}[leftmargin=*]
  \item Beat Coverage Score (BCS): the ratio of aligned musical beats to the generated musical beats ($B_a/B_g$).
  \item Beat Hit Score (BHS): the ratio of aligned musical beats to the ground truth beats ($B_a/B_t$).
  \item F1-score: an integrated assessment of rhythm alignment.
  \item Tempo Difference (TD): average L1 norm of tempo difference between generated and ground truth music.
\end{itemize}
\paragraph{Audio quality}
Frechet Audio Distance (FAD)~\cite{kilgour2019fr} is widely used for evaluating audio quality by measuring the similarity between the generated audio and the ground truth. A lower FAD value indicates more close to the ground truth. We report the FAD based on VGGish audio embedding model~\cite{hershey2017cnn}, which is pre-trained on the YouTube-8M audio dataset~\cite{abu2016youtube}.

\paragraph{Genre similarity}
We employ CLAP~\cite{wu2023large}, a pre-trained large-scale contrastive language-audio model, to compute genre similarity. The CLAP score evaluates the degree of similarity between the CLAP embeddings of the generated audios and the ground truth audios.

\begin{figure*}
\centering
   \includegraphics[width=0.98\linewidth]{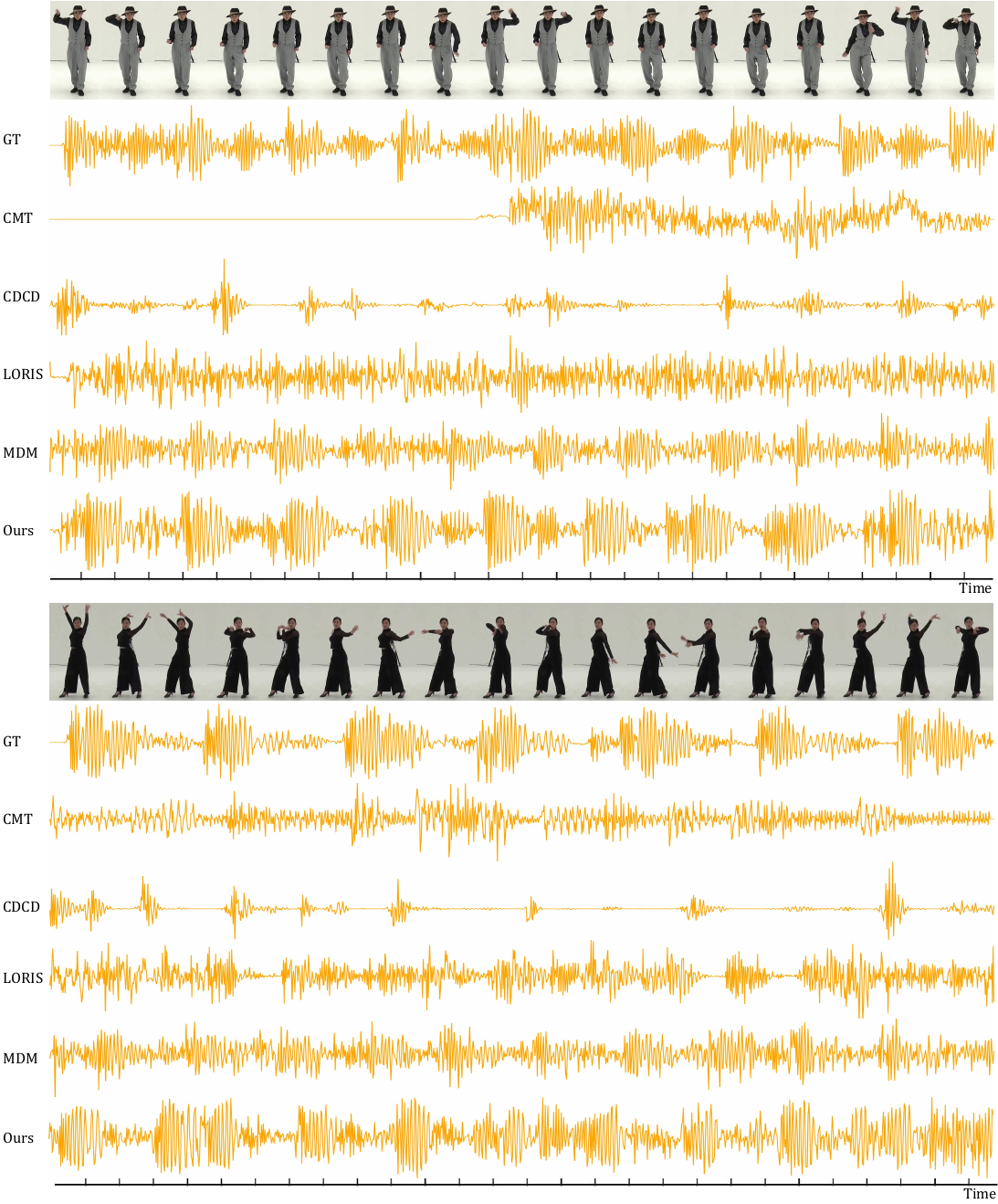}
   \caption{Music visualization examples. ``GT'' and ``Ours'' represents ``Ground Truth'' and ``Ours (MUSICGEN)'', respectively.}
\label{fig:waveform}
\end{figure*}

\subsection{Quantitative Evaluation}
\paragraph{AIST++ dataset~\cite{Li:2021:AIST}}
We conducted experiments on AIST++ dataset by deploying our method on three mainstream base models: the diffusion-based Riffusion~\cite{Forsgren_Martiros_2022}, AudioLDM~\cite{liu2023audioldm} and the autoregressive MUSICGEN~\cite{copet2023simple}. Additionally, we compared our approach against four state-of-the-arts methods: CMT~\cite{di2021video}, CDCD~\cite{zhu2022discrete}, LORIS~\cite{yu2023long}, and MDM~\cite{tan2023motion}. 
We re-implemented CDCD and LORIS and evaluated CMT and MDM using the models provided by the original authors.
Our experimental results on three base models and their comparisons with other methods are shown in Table~\ref{tab:contrast}.
Our method outperforms others across all metrics.
Specifically, Riffusion-based experiments achieve optimal results on all objective metrics except TD, while MUSICGEN-based experiments achieve the best results in subjective metrics. The utilization of large-scale pre-trained models leads to longer inference times.
We additionally compare the phase-aligned versions of each output (see Table~\ref{tab:contrast_align_phase}), and our method achieves the optimal overall rhythm alignment (F1-score), indicating that our method achieves local adaptation. The slight weakness of BCS might be due to CMT, CDCD and LORIS tending to generate fewer beats. While global alignment easily improves their beat precision (BCS), the recall rate (BHS) still shows a significant weakness. In terms of genre similarity and audio quality, our method maintains an obvious advantage. The rhythm alignment metrics for certain superior methods decline because the global offset disrupts the local alignment of the generated results to some extent.

\begin{figure*}
\centering
   \includegraphics[width=1.0\linewidth]{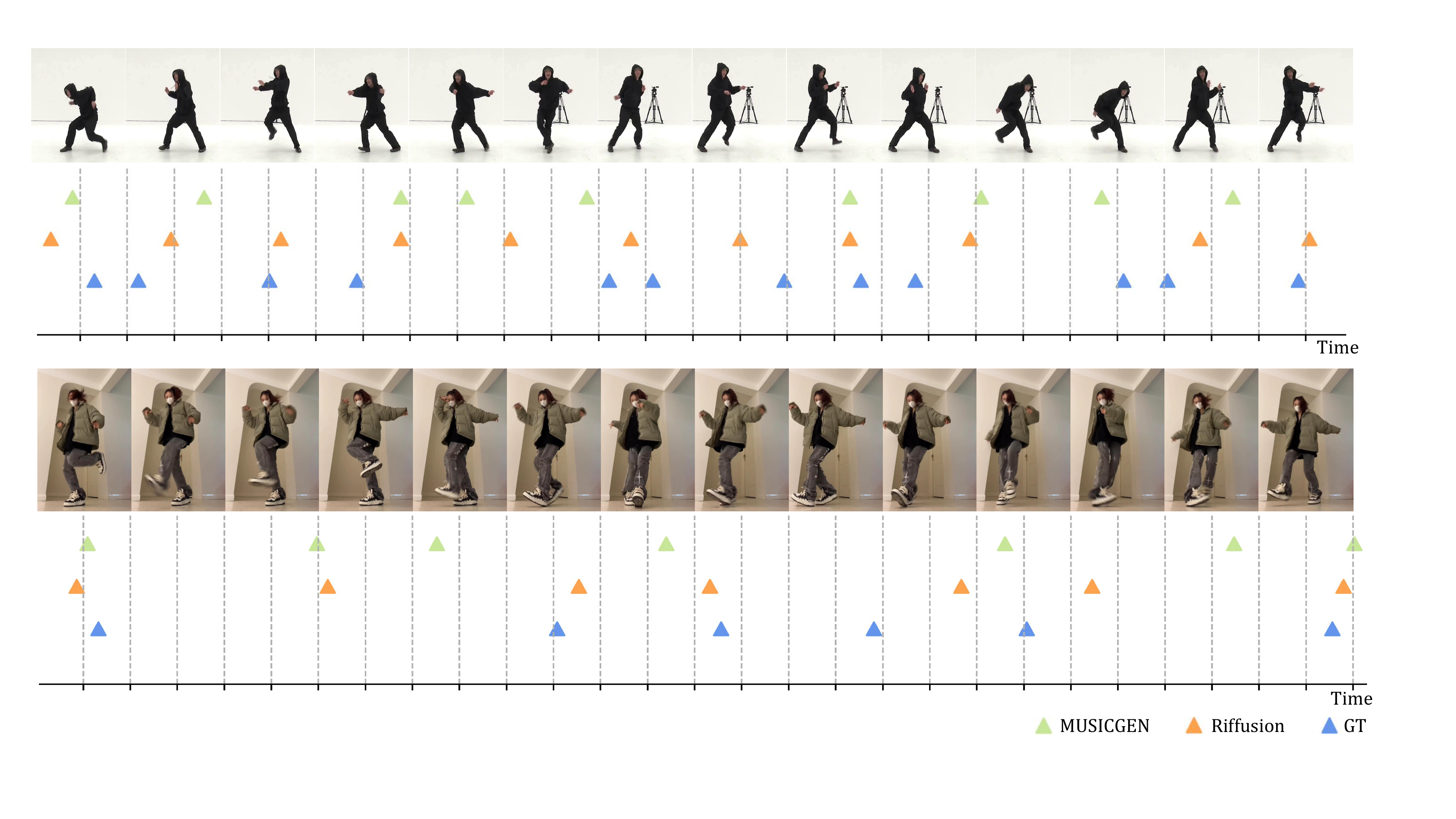}
   \caption{Qualitative examples of beat alignment. \revision{The time scale interval is 0.1s. The distribution of beats show that the majority of the generated beats closely match the ground truth, with offsets below 0.2s. The Riffusion-based model demonstrates slightly better beat alignment compared to MUSICGEN-based model, consistent with quantitative metrics.
   }}
\label{fig:result}
\end{figure*}

\begin{table}
   \caption{Quantitative evaluation on InDV dataset. OM and OR represent Ours (MUSICGEN) and Ours (Riffusion), respectively.}
   \centering
   \resizebox{0.96\linewidth}{!}{
   \begin{tabular}{c||cccc|c|c}
   \hline
  {}&\multicolumn{4}{|c}{Rhythm} & \multicolumn{1}{|c}{Quality} & \multicolumn{1}{|c}{Genre} \\
  \cline{2-7}
     & BCS $\uparrow$  & BHS $\uparrow$ & F1-score  $\uparrow$ & TD $\downarrow$  & FAD $\downarrow$ & CLAP $\uparrow$ \\ 
   \hline
   CMT & 0.3513  & 0.2045  & 0.2585 & 19.80 & 12.65 & \underline{0.4519} \\
   CDCD & 0.3155 & 0.1905  & 0.2375& 22.80 &15.77 & 0.0747\\
   LORIS & 0.2782 & 0.2624  & 0.2701 & 26.15 &13.51 & 0.2975 \\
   MDM & 0.2439 &  0.1985  & 0.2189 & 21.98 &11.30 & 0.4404  \\
   \textbf{OM} & \underline{0.3594} & \underline{0.4012} &  \underline{0.3792} & \underline{18.80} &\underline{5.791} & 0.4048 \\
   \textbf{OR}& \textbf{0.3613}  & \textbf{0.4562} & \textbf{0.4032} & \textbf{12.57}& \textbf{5.633} & \textbf{0.5161} \\
   \hline
   \end{tabular}
   }
   \label{tab:contrast2}
\end{table}

\paragraph{InDV dataset}
Compared to AIST++, our InDV dataset is a more challenging dataset with in-the-wild video settings that contains diverse dance movements. Table~\ref{tab:contrast2} shows the results of the quantitative evaluation of the experiments in the InDV dataset. Our method outperforms other approaches across multiple metrics, which demonstrates the overall robustness of our approach.

\begin{table}
   \caption{Comparison results with music generated at the target tempo.}
   \centering
   \resizebox{0.96\linewidth}{!}{
   \begin{tabular}{c||ccc|c|c}
   \hline
  {}&\multicolumn{3}{|c}{Rhythm} & \multicolumn{1}{|c}{Quality} & \multicolumn{1}{|c}{Genre} \\
  \cline{2-6}
     & BCS $\uparrow$  & BHS $\uparrow$ & F1-score  $\uparrow$  & FAD $\downarrow$ & CLAP $\uparrow$ \\ 
   \hline
   Mustango & 0.3849  & 0.3247  & 0.3522 & 13.33 & 0.4566\\
   Phase-aligned Mustango & 0.3964 & 0.3260  & 0.3578& 14.08 &\underline{0.5230} \\
   \textbf{Ours (MUSICGEN)} & \underline{0.4118}& \underline{0.3874} & \underline{0.3992} & \underline{6.014} &0.4685\\
   \textbf{Ours (Riffusion)}& \textbf{0.4761} & \textbf{0.4398}& \textbf{0.4572} & \textbf{3.416}& \textbf{0.7680}\\
   \hline
   \end{tabular}
   }
   \label{tab:mustango}
\end{table}

\paragraph{Comparison with music generated at the target tempo}
We use Mustango~\cite{melechovsky2023mustango} to generate music at the same tempo as the groundtruth and perform phase alignment.
The comparison results are shown in Table~\ref{tab:mustango}. Our method significantly outperforms the results of Mustango, indicating that our approach provides local adaptation compared to the global control of the tempo.

\paragraph{User study I}
We conduct a Mean Opinion Score (MOS) test. We provide participants with guidelines outlining the evaluation criteria for dance-to-music generation prior to the test. During the test, 82 participants rate a total of 80 samples, including 10 samples from each method and their corresponding ground truth. The evaluations are based on two criteria: dance-music coherence and audio quality.
Dance-music coherence refers to the degree of alignment between the music and the dance video in terms of rhythm and style. Audio quality refers to the musicality of the audio (\eg, noise level, richness of melody). The rating scale ranges from 1 (best) to 5 (worst).
As shown in Table~\ref{tab:contrast}, our approach based on MUSICGEN outperforms other methods in both metrics and achieves an acceptable margin compared to the ground truth. 

\begin{figure*}
\centering
   \includegraphics[width=1.0\linewidth]{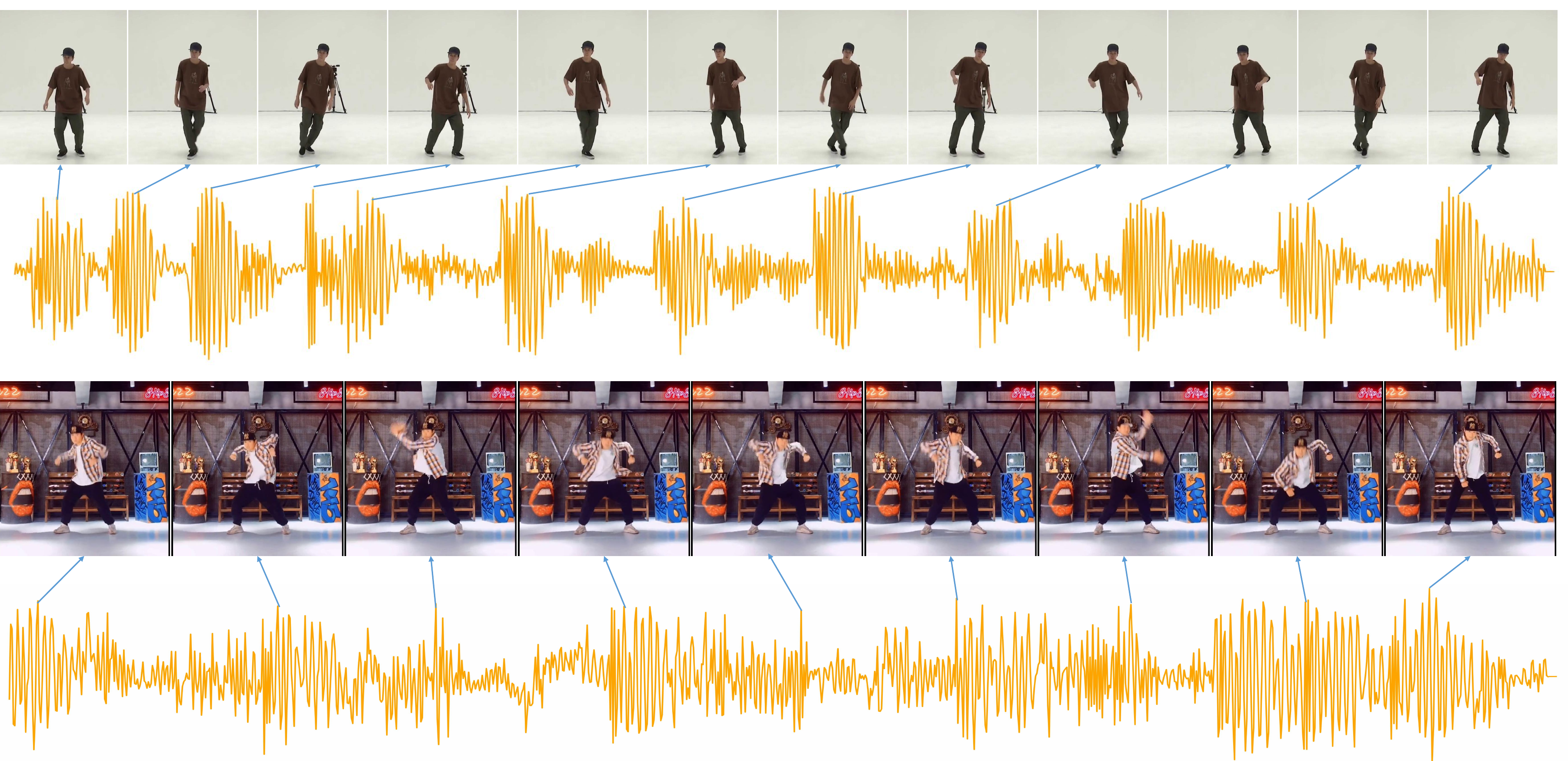}
   \caption{Visualization of results for videos with changes in tempo. The frequency of audio peaks changes in accordance with the variations in video tempo and aligns with dance movements. The first row illustrates an example of slowed-down tempo, while the second row showcases an example of accelerated tempo.}
\label{fig:tempo}
\end{figure*}

\subsection{Qualitative Results}
We visualize the waveforms and corresponding dance movements for both the ground truth and the generated music (see Fig.~\ref{fig:waveform}). Our results show richer dynamics that closely resemble the ground truth, while maintaining alignment with the dance movements. We also compare the beats of the generated music with those of the ground truth (see Fig.~\ref{fig:result}), demonstrating consistent generation of reliable beats with offsets below 0.2. 
The generated audio samples, showcasing the comparison between our method and other approaches, can be accessed on the static webpage provided within the supplementary materials. The samples show that our method generates music with superior coherence and higher quality compared to other approaches.

To verify that our method can adapt to changes in tempo, we manually change the speed of dance videos. The generated audio can still be aligned with the video rhythm while exhibiting tempo variations, as shown in Fig.~\ref{fig:tempo}. Detailed samples can be found in the supplementary video.

Furthermore, samples in our supplementary materials also demonstrate the ability of our method to generate plausible beat-aligned dance music on in-the-wild data (\eg, film clips, virtual human dance videos, and amateur dance videos). Moreover, our method performs well in various other physical activities.
The generated music aligns with the emotional atmosphere depicted in the videos, showcasing the robustness of our approach and its potential for real-world applications.
Benefiting from the editability of the text-to-music model, our method demonstrates the ability to generate diverse music that aligns with the same dance videos.

\subsection{Ablation Study}

\paragraph{Genre encoder}
We first investigate the role of genre encoder that engage into dance-to-music generation. 
We introduce a variant called CLAP that replaces genre labels with audio embeddings derived from other segments of the same audio~\cite{wu2023large}.
When the genre label is replaced with CLAP audio embeddings, experimental results indicate a decrease in all metrics except TD. This observation indicates that the highly repetitive or monotonous audios in the AIST++ dataset is not suitable for facilitating the genre encoder in mapping CLAP audio embeddings to the textual space. As a result, achieving example-based genre control becomes challenging.

\begin{table}
   \caption{Ablation study of our method. 
   ``M'' and ``R'' respectively represents ``MUSICGEN'' and ``Riffusion''. ``GL'' represents genre label.
``CLAP'' represents utilization of CLAP audio embeddings to replace genre labels.
``MLP'' represents the adoption of a Multi-Layer Perceptron (MLP) as the projector in the rhythm encoder.
``Attn+Pos'' denotes the employment of attention mechanism and positional encoding as the projector in the rhythm encoder. }
  \centering
   \resizebox{0.99\linewidth}{!}{
   \begin{tabular}{c|c||cccc|c|c}
   \hline
  \multirow{2}{*}{Base}& \multirow{2}{*}{Encoder} &\multicolumn{4}{|c}{Rhythm} & \multicolumn{1}{|c}{Quality} & \multicolumn{1}{|c}{Genre}\\ 
  \cline{3-8}
    & & BCS $\uparrow$  & BHS $\uparrow$ & F1-score  $\uparrow$ & TD $\downarrow$ & FAD $\downarrow$ &  CLAP $\uparrow$ \\ 
   \hline
   \multirow{3}{*}{M} & CLAP & 0.3925  & 0.3155  & 0.3498 &\textbf{13.47} & 6.821 & 0.4588 \\
   
    & MLP + GL& 0.3958 & 0.3643  & 0.3794 & 15.51 & 6.469 & \textbf{0.4695}\\
 & Attn + Pos & \textbf{0.4118} & \textbf{0.3874}  & \textbf{0.3992} & 16.06 & \textbf{6.014} & 0.4685\\
 \hline
   \multirow{3}{*}{R} & CLAP & 0.4480  & 0.3085  & 0.3654 &\textbf{ 14.49} & 4.611 & 0.6477\\
    & MLP + GL & \textbf{0.4761}  & 
   \textbf{0.4398} & \textbf{0.4572}&20.34 & \textbf{3.416} & ~\textbf{0.7680} \\
 & Attn + Pos & 0.3806 & 0.2768  & 0.3205 & 21.69 & 3.814 & 0.7524\\
   \hline
   \end{tabular}
}
   \label{tab:ablation}
\end{table}

\paragraph{Rhythm encoder}
We design two projectors to map rhythm sequences into the textual space: ``MLP'', which utilizes a combination of multiple linear layers and activation layers as the projector, and ``Attn+Pos'', which incorporates position embedding and attention modules as the projector.
As indicated in Table~\ref{tab:ablation}, experiments using the Riffusion framework demonstrate improved performance with the MLP projector but diminished results with the Attn+Pos projector. In contrast, the experiments conducted with MUSICGEN show enhanced performance with the Attn+Pos projector, suggesting that MUSICGEN is more sensitive to position embeddings.

\subsection{Discussions and Limitations}
Our approach generates diverse and high-quality music that aligns with videos, allowing for personalized and interactive music experiences in the context of dance videos. Furthermore, our method can be extended to in-the-wild data (\eg, ``film'', ``jump rope'' and ``artistic gymnastics'').
For limitation, our method currently supports only fixed-length video segments for music composition. The flexibility of accommodating variable-length segments would enhance the applicability of our method in real-world scenarios.
\section{Conclusion}
This paper introduces an encoder-based textual inversion technique to seamlessly integrate rhythm and genre control into pre-trained text-to-music models. Our approach offers a plug-and-play solution for text-to-music models. We conduct a comprehensive evaluation on two datasets, encompassing rhythm, audio quality, and genre. Experimental results demonstrate that our method can generate high-quality music that is synchronized with the videos across various genres. 
Furthermore, our approach adapts to changes in tempo and enables flexible editing of high-level music attributes using text prompts.
Future research may delve into the integration of multimodal motion information to achieve more robust beat generation results. Additionally, incorporating example-based genre control could provide more flexible user experiences in dance-to-music generation.

\begin{acks}
We thank Xue Song for preparing some dance video data and Minyan Luo for selection of the results. This work was supported in part by the National Natural Science Foundation of China under nos. U20B2070 and 62102162, in part by the Beijing Science and Technology Plan Project under no. Z231100005923033, in part by the National Science and Technology Council under no. 111-2221-E-006-112-MY3, Taiwan, in part by the German Research Foundation (DFG) Project under no. 508324734, and in part by Kuaishou.
\end{acks}

\bibliographystyle{ACM-Reference-Format}
\bibliography{Dance2Music}

\end{document}